\documentclass[10pt,aps,pra,twocolumn,showpacs,superscriptaddress]{revtex4-2}
\usepackage[english]{babel} 	
\usepackage[dvipsnames]{xcolor} 
\usepackage{graphicx} 
\usepackage{bm} 
\usepackage{amsmath} 
\usepackage{amsthm} 
\usepackage{amssymb} 
\usepackage{times} 
\usepackage[pdftex,colorlinks=true,urlcolor= blue,linkcolor=Blue,citecolor=RedViolet]{hyperref}
\usepackage{multirow}
\usepackage{graphicx}
\usepackage{bm}
\usepackage{physics}
\usepackage{ bbold }
\usepackage{comment}

\providecommand{\moy}[1]{\langle #1 \rangle}

\providecommand{\ketbra}[2]{|  #1\rangle \langle #2 |}

\usepackage{soul}

\begin{document}

\title{Anomalous discharging of quantum batteries: the ergotropic Mpemba effect}

\author{Ivan Medina}
\email{ivan.medina@ifsc.usp.br}
\affiliation{Instituto de F\'{i}sica de S\~{a}o Carlos, Universidade de São Paulo, CP 369, 13560-970 São Carlos, Brazil}
\affiliation{School of Physics, Trinity College Dublin, Dublin 2, Ireland}
\author{Oisín Culhane}
\email{oculhane@tcd.ie}
\affiliation{School of Physics, Trinity College Dublin, Dublin 2, Ireland}
\author{Felix C. Binder}
\email{felix.binder@tcd.ie}
\affiliation{School of Physics, Trinity College Dublin, Dublin 2, Ireland}
\affiliation{Trinity Quantum Alliance, Unit 16, Trinity Technology and Enterprise Centre, Pearse Street, Dublin 2, D02YN67, Ireland}
\author{Gabriel T. Landi}
\email{gabriel.landi@rochester.edu}
\affiliation{Department of Physics, University of Rochester, Rochester, New York 14627, USA}
\author{John Goold}
\email{GOOLDJ@tcd.ie}
\affiliation{School of Physics, Trinity College Dublin, Dublin 2, Ireland}
\affiliation{Trinity Quantum Alliance, Unit 16, Trinity Technology and Enterprise Centre, Pearse Street, Dublin 2, D02YN67, Ireland}
\begin{abstract}
Anomalous thermal relaxation is ubiquitous in nonequilibrium statistical mechanics.  An emblematic example of this is the Mpemba effect, where an initially ``hot'' system cools faster than an initially ``cooler'' one. This effect has recently been studied in a variety of different classical and quantum settings. In this Letter, we find a novel signature of the Mpemba effect in the context of quantum batteries. We identify situations where batteries in higher charge states can discharge faster than less charged states. Specifically, we consider a quantum battery encoded in a single bosonic mode that is charged using unitary Gaussian operations. We show that the ergotropy, used here as a dynamical indicator of the energy stored in the battery,  can be recast as a phase space relative entropy between the system's state and the unitarily connected passive state, at each time. Our formalism allows us to compute the ergotropy analytically under dissipative dynamics and allows us to understand the conditions which give rise to a Mpemba effect. We also find situations where two batteries charged to the same value using different operations can discharge at different rates. 
\end{abstract}

\maketitle


 Quantum batteries (QBs) are quantum systems that can be used to conceptually understand work deposition and extraction in quantum thermodynamics~\cite{Alicki2013,Hovhannisyan_2013,Binder2018,QBRev}. Differently from classical batteries, QBs can be manipulated using physical operations that induce quantum correlations. In this regard, the study of QBs aims to determine whether quantum features can provide benefits for energy storage and release. Within this framework we can understand how quantum effects such as the use of entangling operations can be exploited for enhanced charging power~\cite{Binder_2015,Campaioli_2017}, how quantum correlations can connect to work extraction~\cite{Francica_2017} and how coherent contributions play a role~\cite{Francica_2020}. A charged battery is in a nonpassive state with respect to its Hamiltonian. The maximum amount of energy that can be extracted from a nonpassive state via cyclic unitary operations is known as the ergotropy~\cite{ergotropy} and is a state-dependent measure of the recoverable energy from the battery. In reality, physical systems interact with some environment, and through these interactions, part of the energy stored in the QB is leaked to the environment, thus diminishing the charge over time. Various protocols have been developed to stabilize the charge of a quantum battery coupled to an environment. One technique is through the use of dark states --- quantum states that do not relax when coupled to the reservoir. These dark states can be generated through the use of decoherence-free subspaces~\cite{Zanardi_1997,Liu_2019} or spontaneously created through interactions between the charger and reservoir~\cite{Quach_2020}. The ergotropy can also be stabilized through measurements of the battery~\cite{Gherardini2020,malavazi2024}. While the majority of works in QBs deal with discrete systems, recent works have extended the framework to continuous variable (CV) settings~\cite{Brown_2016,PhysRevA.100.042104,CVQB,CVQB2}.

\begin{figure}[t]
	\centering
	\includegraphics[width=0.48\textwidth]{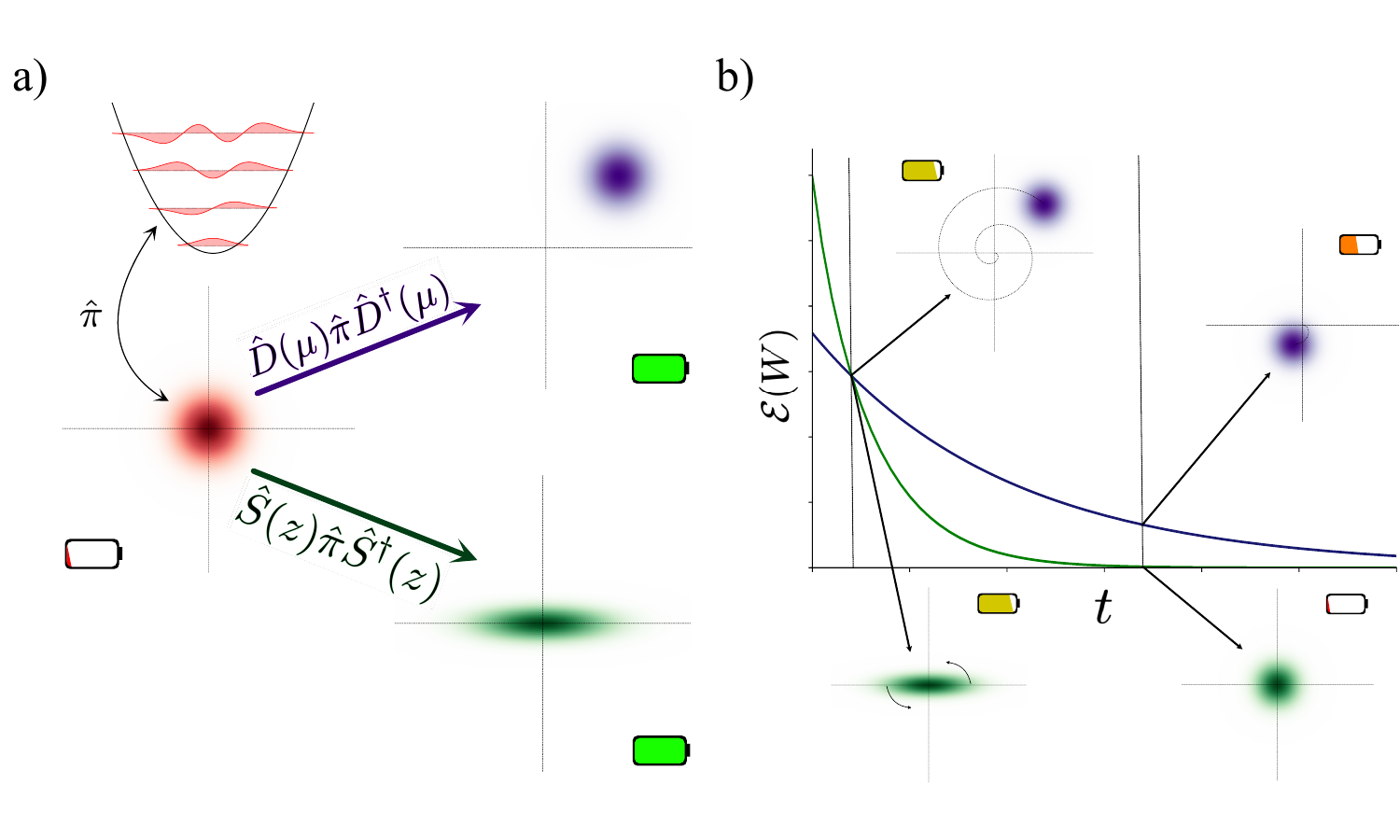}
	\caption{(a) Single mode thermal-state quantum oscillator with no stored ergotropy can be charged via squeezing and displacement operations, preserving the system's Gaussianity. (b) When weakly coupled to a thermal bath, the charge is lost to the environment. Charge generated via squeezing operations is lost faster than charge stored via displacement operations, leading to an ergotropic Mpemba effect. }
	\label{fig1}
\end{figure}

In this Letter, we provide a fundamental contribution to the theory of quantum batteries in the Gaussian CV case by describing how to slow down the discharge rate through the quantum Mpemba effect. Our results are obtained through an analytic treatment by virtue of the Gaussianity preserving dynamics studied. The Mpemba effect, commonly described as hot water freezing faster than cold water, was discussed in the 1960s by Mpemba and Osbourne~\cite{Mpemba1969} and independently by Kell~\cite{Kell_1969}. The phenomenon has been explored using Markovian microscopic systems~\cite{Raz2017,Klich_2019} and has been experimentally demonstrated in colloidal systems~\cite{Kumar_2022} and trapped ion platforms~\cite{Shapira_2024,Zhang_2025}. One can also find the effect in open quantum system theory~\cite{Moroder_2024,Carollo_2021,Carollo_2022,Nava_2019,Chatterjee_2023}, and it has been investigated in many-body dynamics in connection with symmetry restoration~\cite{Ares_2023, Turkeshi_2024}. Recently, the Mpemba effect has been extended to bosonic systems~\cite{longhi20241,longhi20242,furtado2024} where it was related to nonclassical states of light and squeezed reservoirs. In this Letter, we study the Mpemba effect through the lens of work extraction and demonstrate that it can be used to slow down the loss of energy that can be unitarily extracted from the battery, quantified through the ergotropy. In other words some batteries with higher charge will discharge faster than batteries with lower charge. This is illustrated in Fig. \ref{fig1}, where it is shown that an ergotropic Mpemba effect can occur depending on the unitary Gaussian operation used to charge the QB.

\paragraph*{Ergotropy for Gaussian states -} Given a system described by a Hamiltonian $\hat{H}$ and a density matrix $\hat{\rho}$, the {\it ergotropy}~\cite{ergotropy,passive,Alicki2013} is the maximum amount of work that can be extracted from a nonpassive state by means of a cyclic unitary operation. A cyclic unitary operation can be implemented by a cyclic variation of the Hamiltonian  through the control field $V(t)$ in a fixed time interval $t\in[t_1,t_2]$, such that $H(t)=H+V(t)$ and $V(t_1)=V(t_2)=0$. The ergotropy is defined by
\begin{align}
	\mathcal{E}(\hat{\rho}, \hat{H})=E(\hat{\rho})-E(\hat{\pi}),\label{erg}
\end{align}
where $E(\hat{\rho})={\rm Tr}\{\hat{H} \hat{\rho}\}$ and $\hat{\pi}$ is the passive state associated to $\hat{\rho}$, i.e., the state obtained after the maximum amount of work is unitarily extracted from $\hat{\rho}$. The energy of this state can be obtained from the optimization $E(\hat{\pi})=\min_{\hat{U}}{\rm Tr}\{\hat{H} \hat{U} \hat{\rho} \hat{U}^\dagger\}$. Mathematically, we can understand the passive state as follows. Let $\hat{\pi}=\sum_k p_k \ketbra{p_k}{p_k}$ with $p_{k}\geq p_{k+1}$  and $\hat{H}=\sum_k\epsilon_k\ketbra{\epsilon_k}{\epsilon_k}$ with $\epsilon_k\leq\epsilon_{k+1}$. The condition for the passivity of $\hat{\pi}$ with respect to $\hat{H}$ is that $[\hat{H},\hat{\pi}]=0$ and $p_n\geq p_m$ when $\epsilon_n<\epsilon_m$. 

Our quantum battery is a single bosonic mode described by the Hamiltonian ($\hbar=1$, $k_B=1$) $\hat{H}=\omega(\hat{a}^\dagger \hat{a}+1/2)$ with creation (annihilation) operator $\hat{a}$ ($\hat{a}^\dagger$) satisfying the canonical commutation relation $[\hat{a},\hat{a}^\dagger]=1$. We also assume that our system is restricted to the class of Gaussian states, which are completely characterized by the mean vector $\vec{v}=(\moy{\hat{a}}, \moy{\hat{a}^\dagger})$ and the covariance matrix $	\Theta_{ij}=\frac{1}{2}\moy{\{u_i,u_j^\dagger\}}-\moy{u_i}\moy{u_j^\dagger}$, with $\vec{u}=(\hat{a}, \hat{a}^\dagger)$. An arbitrary Gaussian  state $\hat{\rho}$ can  be represented in phase space by the Wigner function as
\begin{align}
W(\vec{\alpha})=\frac{1}{\pi \sqrt{|\Theta|}}\exp\left[-\frac{1}{2}(\vec{\alpha}-\vec{v})^\dagger\Theta^{-1}(\vec{\alpha}-\vec{v})\right],\label{wcov}
\end{align}
where $|\Theta|=\det{\Theta}$ and $\vec{\alpha}=(\alpha,\alpha^*)$. In this work, we use the notation $\text{d}\vec{\alpha}$ to indicate integration over the phase space, i.e.,  $\text{d}\vec{\alpha}=\text{d}{\rm Re}[\alpha]\, \text{d}{\rm Im}[\alpha]$, where Re$[\alpha]$ and Im$[\alpha]$ denote the real and imaginary parts of $\alpha$, respectively. From the Wigner function, it is possible to compute any observable quantity. In particular, the mean energy of the system is given by
\begin{align}
	E(W)=\omega\int d\vec{\alpha} |\alpha|^2 W(\vec{\alpha}).\label{energy}
\end{align}

Our first result connects the ergotropy of a bosonic mode with the relative Wigner entropy~\cite{PhysRevLett.109.190502},
\begin{align}
	K[W_1||W_2]=\int d\vec{\alpha} W_1(\vec{\alpha}) \ln \frac{W_1(\vec{\alpha})}{W_2(\vec{\alpha})},\label{relatw}
\end{align}
between any two Gaussian Wigner functions $W_1(\vec{\alpha})$ and $W_2(\vec{\alpha})$. By using Eqs. \eqref{erg}, \eqref{energy}  and \eqref{relatw} we show in the Appendices  \ref{AppA} and \ref{AppaB} that the ergotropy can be written as
\begin{align}
	\mathcal{E}(W)=\omega f(\beta_\pi)K[W||W_{\rm \pi}],\label{werg4}
\end{align}
where $W_{\rm \pi}(\vec{\alpha})$ is the Wigner function associated to the thermal passive state $\hat{\pi}$ with inverse temperature $\beta_\pi$ defined as $f(\beta_\pi)=\bar{n}_\pi+1/2$ where $\bar{n}_\pi=(e^{\beta_\pi \omega}-1)^{-1}$. The function $f(\beta)$ is related to the covariance matrix of a thermal state with inverse temperature $\beta$, which is given by $\Theta_{\rm th}=f(\beta)\mathbb{I}$. We use the notation with the subscript $\pi$ when referring to the thermal state that is the passive state associated to $\rho$. As $W_{\pi}(\vec{\alpha})$ is the passive state associated to $W(\vec{\alpha})$, they are unitarily connected, and hence must have the same entropy, i.e.,	$S(W)=S(W_{\pi})$. For Gaussian states, we can use the Wigner entropy
\begin{align}
\label{Eq: Wigner Entropy}
S(W)=-\int d \vec{\alpha} W(\vec{\alpha})\ln W(\vec{\alpha}),
\end{align}
which is directly related to the Reyni-2 entropy of the quantum state \cite{PhysRevA.99.052104,PhysRevLett.109.190502,PhysRevLett.118.220601,PhysRevLett.121.160604}. This entropic equality implies the following constraint for the temperature of the passive state (see Appendix \ref{AppaB})
\begin{align}
f(\beta_\pi)=\sqrt{|\Theta|},\label{constraint}
\end{align}
which has the following important consequence: for Gaussian states the energy of the passive state (associated with the nonpassive battery state $\hat{\rho}$), $E(W_\pi)=\omega \sqrt{|\Theta|}$, is determined only by the covariance matrix of the nonpassive state $\hat{\rho}$.

\paragraph*{Gaussianity preserving dynamics and ergotropy rate -} To account for a more realistic scenario, we assume our bosonic system to be weakly coupled to a reservoir at inverse temperature $\beta$, which causes the system to evolve according to the following Markovian master equation:
\begin{align}
	\dot{\hat{\rho}}=-i[\hat{H},\hat{\rho}]+\gamma(1+\bar{n})\mathcal{D}_{\hat{\rho}}[\hat{a}]+\gamma\bar{n}\mathcal{D}_{\hat{\rho}}[\hat{a}^\dagger],\label{MEQ}
\end{align}
where $\gamma$ is the dissipation rate and $\mathcal{D}_{\hat{\rho}}[\hat{L}]=\hat{L} \hat{\rho} \hat{L}^\dagger -\frac{1}{2} (\hat{L}^\dagger \hat{L} \hat{\rho}+\hat{\rho} \hat{L}^\dagger \hat{L} )$ is the dissipator describing the nonunitary part of the dynamics. The parameter $\bar{n}=f(\beta)-1/2$ is related to the fixed point of Eq.~\eqref{MEQ}, which is the thermal state $\hat{\rho}_{\rm th}=(1-e^{-\beta \omega})e^{-\beta \omega \hat{a}^\dagger \hat{a}}$.
Since the system Hamiltonian is quadratic and the Lindblad generators are linear in  $\hat{a}$ and $\hat{a}^\dagger$, the dynamics is Gaussianity preserving \cite{PhysRevA.106.052206,PRXQuantum.5.020201}. The system's Wigner function will preserve the form of Eq.~\eqref{wcov}, but $\Theta$ and $\vec{v}$ will acquire a time-dependent component. While the mean vector evolution is trivially given by $\vec{v}=(\moy{\hat{a}}_0 e^{-i \omega t},\moy{\hat{a}^\dagger}_0 e^{+i \omega t})e^{-\frac{\gamma}{2}t}$, the covariance matrix $\Theta$ obeys a Lyapunov equation \cite{PRXQuantum.5.020201}
\begin{align}
	&\dot{\Theta}=\Lambda \Theta+\Theta \Lambda^\dagger + F,\label{lyapunov}\\
	&\Lambda=-\frac{1}{2}\begin{bmatrix}
		\gamma+2i\omega & 0 \\
		0 & \gamma-2i\omega 
	\end{bmatrix}, \ \ 
	F=f(\beta)\hat{\mathbb{I}}, \nonumber
\end{align}
where $\moy{\cdot}_0$ denotes the average value in the initial state and $\hat{\mathbb{I}}$ is the dimension-2 identity matrix. The Lyapunov equation [Eq.~\eqref{lyapunov}] can be analytically solved, as detailed in the Appendix \ref{AppaD}.

Because the system undergoes dissipative dynamics, the battery will lose charge, which is reflected in the depletion of ergotropy over time.  By taking the time derivative of the ergotropy, we find that the ergotropic loss rate is given by
\begin{align}
\dot{\mathcal{E}}(W)=-\Phi(W)-\omega \sqrt{|\Theta|}\dot{S}(W), \label{wergdtime2}
\end{align}
where we defined the energy flux as $\Phi(W)=-\dot{E}(W)$. Notice that the rate depends not only on the energy relaxation but also on the entropy change in the system. Interestingly, because of the unitary constraint between $W(\vec{\alpha})$ and $W_{\rm \pi}(\vec{\alpha})$, the energy flux of the passive state, $\Phi(W_{\rm \pi})$, is connected to the derivative of $S(W)$ by $\Phi(W_{\rm \pi})=	-\omega\sqrt{|\Theta|}\dot{S}(W).$ The relation established in Eq. \eqref{wergdtime2} unravels an important aspect of dissipative Gaussian QBs. The first term is a straightforward energy dissipation of the battery, and it is related to the energy that can be recovered by unitary charging. The second term, however, is proportional to the change of the system's entropy, and it dictates how the passive state energy rate changes with respect to the spectrum of the system's state. This change in the spectrum of the state represents the contribution of the energy that is lost to the environment and cannot be recovered by means of unitary operations, i.e.,  it represents a depletion on the QB ``quality". It is important to mention that we are assuming that one is able to perform the unitary operations necessary to extract work in a time scale that is much shorter than the dissipative dynamics.  
 \begin{figure}[t]
	\centering
	\includegraphics[width=0.48\textwidth]{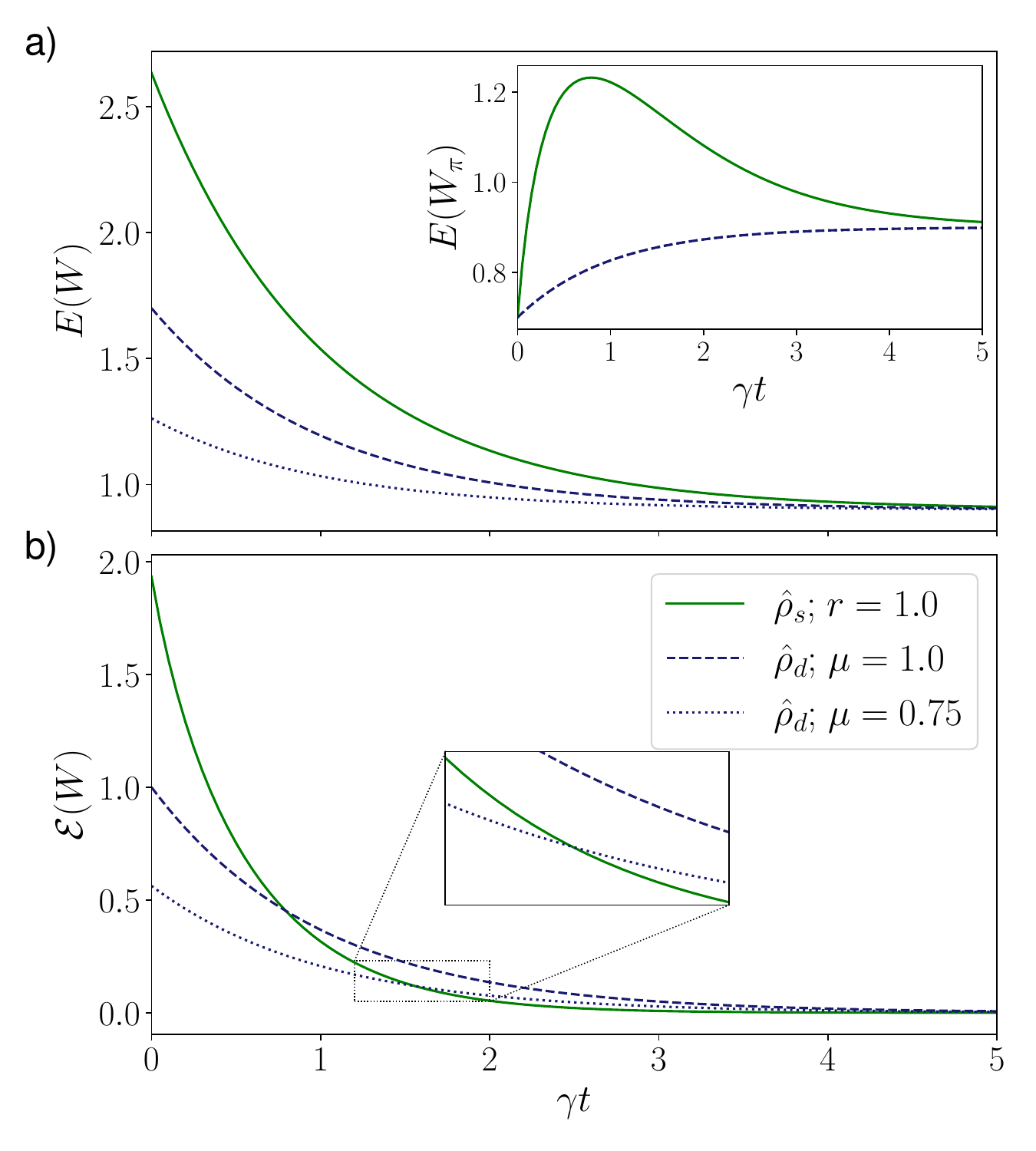}
	\caption{Energetic and ergotropic time evolution of a squeezed (green) and a displaced (blue) single Bosonic mode. (a) The energy evolution of the oscillator displays no crossing between the squeezed and displaced states. {\it Inset}: Evolution of the associated passive state of the oscillator; the passive state associated with the squeezed state is nonmonotonic in time. (b) Ergotropic evolution of the oscillator. As the ergotropy is the difference between the energy of the state and the associated passive state, the behavior of the passive energies generates a Mpemba effect, where despite having a higher initial ergotropy, the squeezed state relaxes much faster than the displaced state. The relevant parameters were set to $\bar{n}_\pi=0.2$ and $\bar{n}=0.4$.}
	\label{fig2}
\end{figure}

\begin{figure}[t]
	\centering
\includegraphics[width=0.48\textwidth]{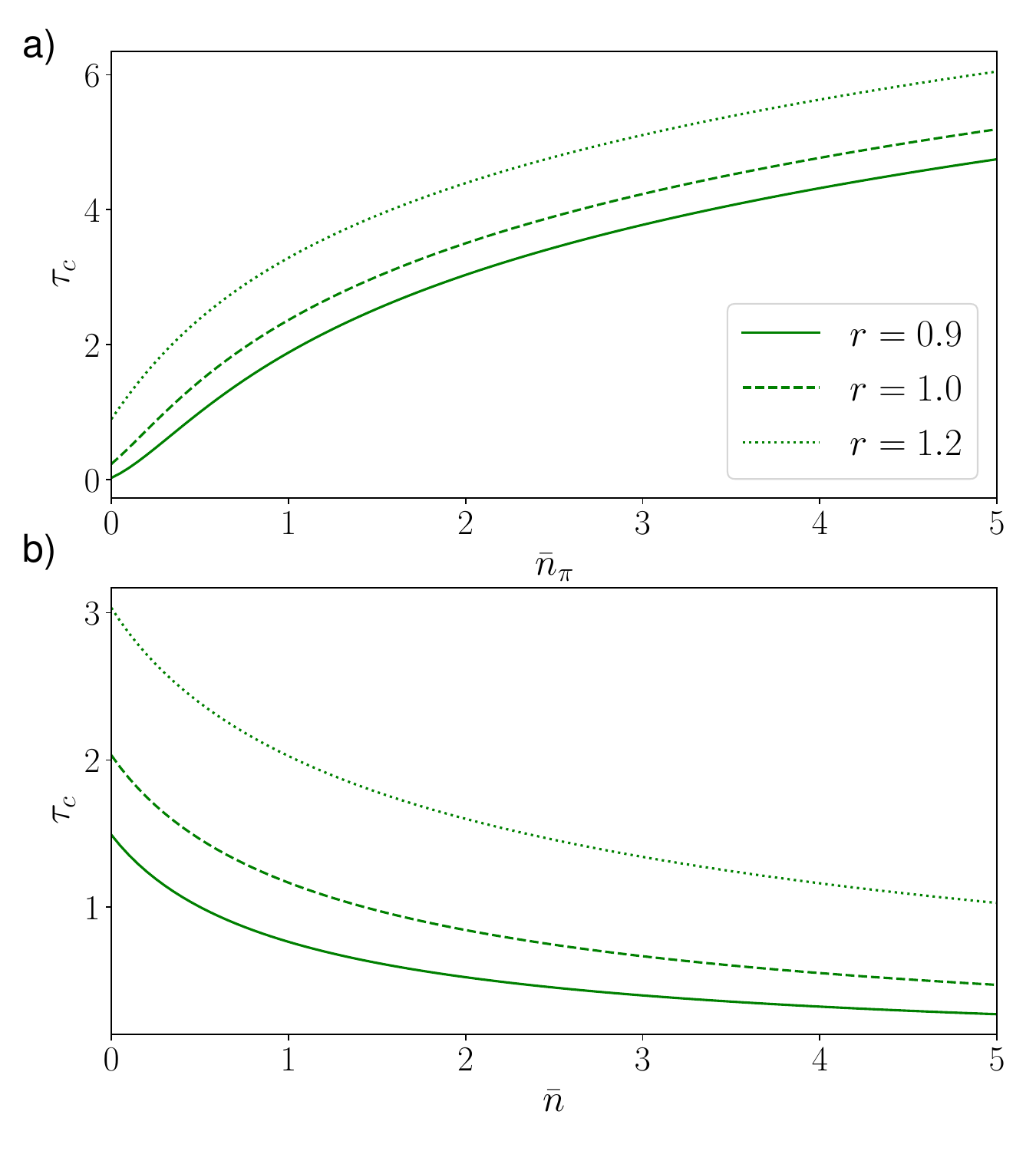}
	\caption{Time, $\tau_c$, for the ergotropic Mpemba crossing to occur as a function of environment and initial passive state temperatures for different values of squeezing parameter $r$. (a) We fixed $\bar{n}=0.5$ while varying $\tau_c$  with $\bar{n}_\pi$. (b)  We fixed $\bar{n}_\pi=0.5$ while varying $\tau_c$  with $\bar{n}$. For all plots, $\mu=1.0$ is fixed. }
	\label{fig3}
\end{figure}

\paragraph*{Displacement and squeezing contributions -} Working with Eq.~\eqref{werg4} and explicitly taking into account the constraint [Eq. ~\eqref{constraint}], we can split the ergotropy as  $\mathcal{E}(W)=\mathcal{E}_{\rm d}(\vec{v})+\mathcal{E}_{\rm s}(\Theta)$ (see Appendix \ref{AppaC}), where
$\mathcal{E}_{\rm d}(\vec{v})$ is the  contribution coming from the mean vector and $\mathcal{E}_{\rm s}(\Theta)$ is the  contribution coming from the covariance matrix.
The most general single-mode Gaussian state is generated by a squeezed displaced thermal state of the form $\hat{\rho}=\hat{S}(z)\hat{D}(\mu) \hat{\pi}\hat{D}^\dagger(\mu)\hat{S}^\dagger(z)$,
where $\hat{D}(\mu)=\exp[\mu \hat{a}^\dagger-\mu^*\hat{a}]$ and $\hat{S}(z)=\exp[\frac{1}{2}(z^*\hat{a}^2-z\hat{a}^{\dagger 2})]$ are the displacement and squeezing operators, respectively \cite{PKnight}. 
While the displacement operator only affects the mean vector, the squeezing operator only affects the covariance matrix. Thus, $\mathcal{E}_{\rm d}(\vec{v})=\mathcal{E}_{\rm d}(\mu)$ will depend only on $\hat{D}(\mu)$ and $\mathcal{E}_{\rm s}(\Theta)=\mathcal{E}_{\rm s}(r)$ will depend only on $\hat{S}(z)$, allowing us to write (see Appendix \ref{AppaE})
\begin{align}
	&\mathcal{E}_{\rm d}(\mu)=\omega |\mu|^2,\label{displaceergo}\\
	&\mathcal{E}_{\rm s}(r)=\omega f(\beta_\pi)[\cosh(2 r)-1],\label{squezergo}
\end{align}
where $r=|z|$ is the squeezing parameter. The Gaussianity preserving dynamics [Eq. \eqref{MEQ}] does not couple the first and second moments, so the displacement and squeezing contributions evolve independently in time, keeping the same shape as Eqs. \eqref{displaceergo} and \eqref{squezergo}. However, $\mu$, $r$, and $\beta_\pi$ will depend on time. The displacement parameter will have an exponential dependence $\mu_t=\mu e^{-(i\omega+\gamma/2)t}$ and $r$ and $\beta_\pi$ will have a nontrivial dependence on $t$, given implicitly by the functions
\begin{align}
	& f(\beta_t)=\sqrt{\Delta_\beta^2+4f(\beta_\pi)f(\beta)e^{-2\gamma t}(e^{\gamma t}-1)\sinh^2(r)},\label{betatime}\\
	& \cosh(2 r_t)=\frac{1}{f(\beta_t)}[\Delta_\beta+2f(\beta_\pi)e^{-\gamma t}\sinh^2(r)],
\end{align}
where $\Delta_\beta=[f(\beta_\pi)-f(\beta)]e^{-\gamma t}+f(\beta)$. We remark on an important distinction between the displacement and squeezing contributions.  On the one hand, since the displacement operator does not affect $\Theta$, the thermal contribution of its energy is not affected by the displacement parameter $\mu$. Consequently, the thermal part of the energy of the displaced thermal state cancels with the energy of the passive state, and the ergotopic contribution $\mathcal{E}_{\rm d}(\mu)$ is independent of $\beta_\pi$. In fact, the energy of the passive state associated with a displaced thermal state evolves in time as $\omega\Delta_\beta$, which is exactly the solution for a thermal state evolving according to Eq. \eqref{MEQ}. On the other hand, the squeezing parameter affects the thermal contribution [see Eq.~\eqref{betatime}] of the thermal squeezed state, and as a consequence, the ergotropy will depend on both the temperature $\beta_\pi$ of the initial passive state and the equilibrium temperature $\beta$. In this case, the energy of the passive state evolves as $\omega f(\beta_t)$, with $f(\beta_t)$ given by Eq.~\eqref{betatime}, where we can see that there is a positive contribution to the thermal energy coming from the squeezing parameter $r$, through the nonlinear function $\sinh^2(r)$. This difference in the passive state evolution will be crucial to understand the anomalous relaxation of the ergotropy, as we will discuss in what follows.
\begin{figure}[t!]
	\centering
	\includegraphics[width=0.48\textwidth]{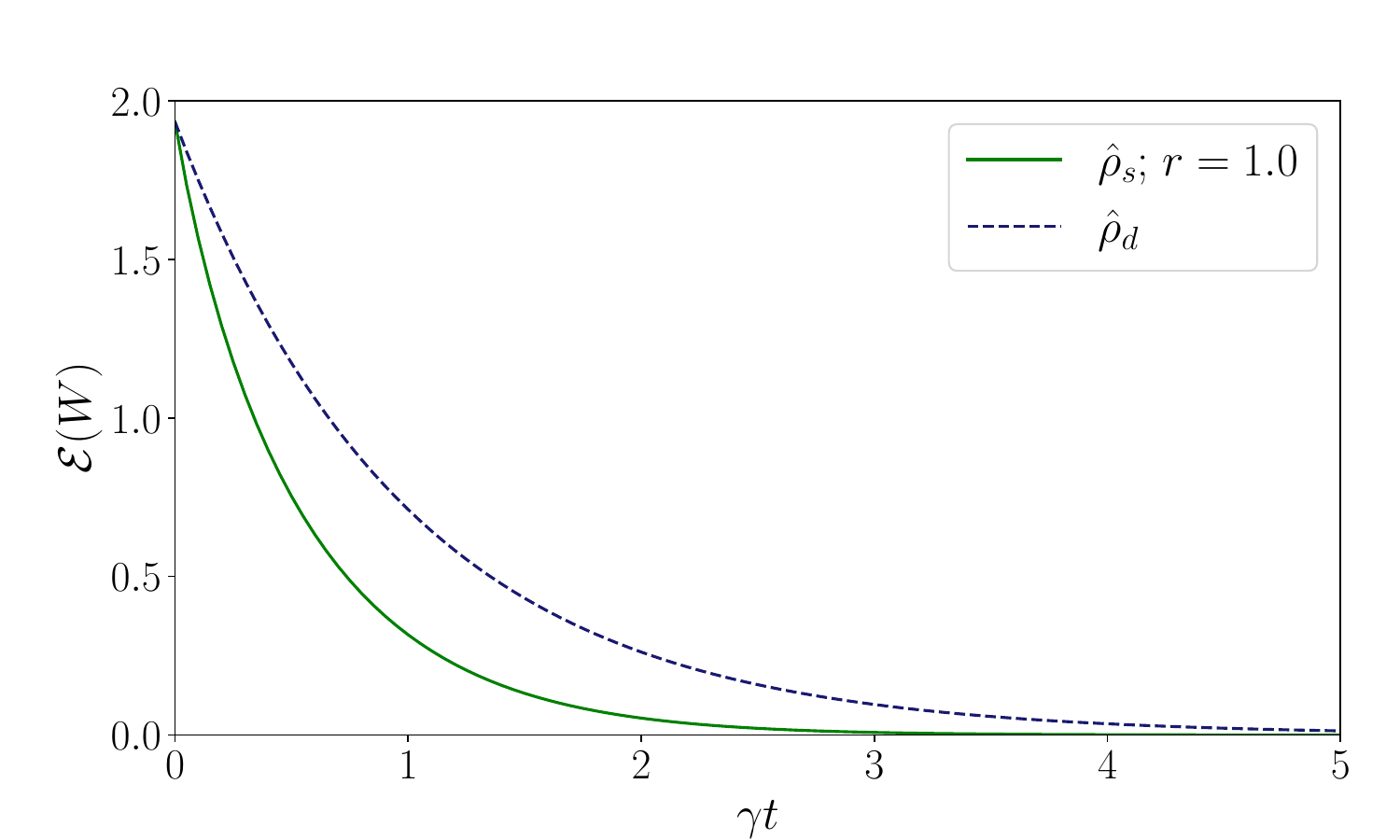}
	\caption{Ergotropy as a function of time for a squeezed thermal state (green line) and a displaced thermal state (blue-dashed line). Following the formula of Eq. \eqref{fastdisc}, we see that even though the states initially have the same ergotropy, the squeezed thermal state loses its charge faster. $\bar{n}_\pi=0.2$ and $\bar{n}=0.4$ were selected as parameters. }
	\label{fig4}
\end{figure}
\paragraph*{Anomalous discharging effect -}  The relaxation of the ergotropy is described by a competition between the relaxation of the state's energy and the energy of its associated passive state. For both displaced and squeezed thermal states their energies relax exponentially. However, as the energy of the passive state associated with the squeezed thermal state has an enhancement coming from the parameter $r$, we always have that $\omega f(\beta_t)\geq\omega \Delta_\beta$; consequently, the ergotropy will always have a higher contribution coming from its passive state, making it relax faster than the ergotropy for a displaced thermal state. To highlight this in Fig.~\ref{fig2} (a), we show the energies $E(W)$ for a squeezed thermal state, $\hat{\rho}_{ s}=\hat{S}(z)\hat{\pi}\hat{S}^\dagger(z)$, with $r=1$ (green line)  and for displaced thermal states, $\hat{\rho}_{d}=\hat{D}(\mu)\hat{\pi}\hat{D}^\dagger(\mu)$, with $\alpha=0.75$ (blue-dashed line) and $\alpha=1.0$ (blue-solid line). For all cases, one can see that the states' energies $E(W)$ relax at the same rate --- there are no crossings in the evolution. We show the associated passive state energies in the inset of Fig.~\ref{fig2} (a). For displaced thermal states, the associated passive energy relaxes monotonically towards the equilibrium state energy of $\omega f(\beta)=0.9$. The passive energy associated with the squeezed state shows nonmonotonic behavior; it increases until it reaches a maximum around $E(W_\pi)\approx 1.2$, before relaxing towards the equilibrium state energy. Now, examining the ergtoropy in Fig~\ref{fig2} (b), we observe an ergotropic Mpemba effect --- the squeezed state loses its charge faster than the displaced states with initially lower charges. 
 
Interestingly, we can find an analytical expression for the time $t_c$ at which the ergotropic Mpemba crossing occurs. Considering the case of an initially squeezed thermal state with higher ergotropy than a displaced thermal state, i.e., $\mathcal{E}_{t=0}(r)>\mathcal{E}_{t=0}(\mu)$, the time $t_c$ at which a crossing occurs, $\mathcal{E}_{t=t_c}(r)=\mathcal{E}_{t=t_c}(\mu)$, is
\begin{align}
\tau_c=\ln{\left[1+\frac{[|\mu|^2-2\cosh^2(r)f(\beta_\pi)]}{2|\mu|^2f(\beta)[|\mu|^2-2\sinh^2(r)f(\beta_\pi)]^{-1}}\right]},
\end{align}
where $\tau_c=\gamma t_c$. In Fig.~\ref{fig3}, we show the time $\tau_c$ as a function of the temperatures $\beta_\pi$ and $\beta$ , through $\bar{n}_\pi$ and $\bar{n}$, for different values of $r$ while keeping $\mu=1.0$ fixed. We see that the higher the temperature $\beta_\pi$ is, the longer it takes for the Mpemba effect to occur. This happens because the ergotropy for the squeezed thermal state increases with $f(\beta_\pi)$ [see Eq.~\eqref{squezergo}], while the ergotropy for the displaced thermal state does not depend on $f(\beta_\pi)$. This increases the initial difference between the charges of the two states, pushing the Mpemba effect to occur at larger times. On the other hand, we see that the higher the temperature of the environment is, the faster the Mpemba effect occurs. In this case, this happens because the temperature $\beta$ increases the contribution of the passive state associated with the squeezed thermal state, making the ergotroy depletion faster. 

Finally, this difference in the ergotropy relaxation for the displaced and squeezed thermal states gives rise to the following novel effect.  For  given parameters $r$ and $\beta_\pi$, we can always find an amplitude $\mu$ which satisfies $\mathcal{E}_{t=0}(\mu)=\mathcal{E}_{t=0}(r)$; namely
\begin{align}
\mu=\pm\sqrt{f(\beta_\pi)[\cosh(2r)-1]}.\label{fastdisc}
\end{align}
Therefore, this allows us to conclude that charging a battery with a displacement operation is a better resource than charging one using a squeezing operation since using the first will make the ergotropy dissipate slower. This is illustrated in Fig.~\ref{fig4}.

\paragraph*{Conclusion -}
Work extraction from quantum systems is a central focus in the field of quantum thermodynamics. We have shown that for Gaussian systems the ergotropy (maximal extractable work from nonpassive state) can be written in terms of a relative entropy which can be used to derive analytic results. In this Letter, we show that a Mpemba effect occurs in the discharge dynamics of quantum batteries. Furthermore, our work represents a rare example of Mpemba physics, which is analytically tractable and sheds important quantitative insight into how the effect occurs. Perhaps most importantly, by formulating the problem in the phase space representation, we open the door to immediate experimental investigation, as there are a multitude of experimental platforms in quantum optics and mesoscopic physics where phase-space distributions are routinely extracted.  

\begin{acknowledgments}
\noindent 

I.M. acknowledges financial support from São Paulo Research Foundation - FAPESP (Grants No. 2022/08786-2 and No. 2023/14488-7). O.C. is supported by the Irish Research Council under
grant number 210500. F.C.B. acknowledges funding from the Irish Research Council (grant IRCLA/2022/3922) and the John Templeton Foundation (grant 62423). J.G. is supported by a SFI- Royal Society University Research Fellowship and funding from Research Ireland under the QUAMNESS grant which is an EPRSC joint funding initiative. 
\end{acknowledgments}

\bibliographystyle{unsrt}
\bibliography{refs.bib}

\onecolumngrid

\appendix

\section{Entropy and Relative Entropy of a Gaussian Distribution}\label{AppA}
In this appendix, we derive general expressions for the Wigner entropy [Eq. \eqref{Eq: Wigner Entropy}] and the relative Wigner entropy [Eq. \eqref{relatw}] for Gaussian states in terms of the vector of mean values, $\vec{v} = \left(\langle\hat{a}\rangle,\langle\hat{a}^\dagger\rangle\right)$, and the covariance matrix
\begin{equation}
	\Theta = 
	\begin{pmatrix}
		\frac{1}{2}\left(\langle\hat{a}^\dagger\hat{a}\rangle + \langle\hat{a}\hat{a}^\dagger\rangle\right)- \langle\hat{a}^\dagger\rangle\langle\hat{a}\rangle & \langle\hat{a}^2\rangle - \langle\hat{a}\rangle^2\\
		\langle\left(\hat{a}^\dagger\right)^2\rangle - \langle\hat{a}^\dagger\rangle^2 & \frac{1}{2}\left(\langle\hat{a}^\dagger\hat{a}\rangle + \langle\hat{a}\hat{a}^\dagger\rangle\right)- \langle\hat{a}^\dagger\rangle\langle\hat{a}\rangle
	\end{pmatrix},\label{covmatrix}
\end{equation}
where $\vec{\alpha} = \left(\alpha,\alpha^*\right)$. An arbitrary Gaussian state $\hat{\rho}$ can be expressed in the phase space representation by the Wigner function as
\begin{equation}
	W(\vec{\alpha})=\frac{1}{\pi \sqrt{|\Theta|}}\exp\left[-\frac{1}{2}(\vec{\alpha}-\vec{v})^\dagger\Theta^{-1}(\vec{\alpha}-\vec{v})\right],
\end{equation}
where $|\Theta|=\text{det} \{\Theta\}$ denotes the determinant of the covarinace matrix. The Wigner entropy can be seen as an expectation value, i.e.
\begin{equation}
	\label{Eq: Wigner_Entropy_Expectation}
	S(W)=-\int d \vec{\alpha} W(\vec{\alpha})\ln W(\vec{\alpha}) = -\mathbb{E}_{W}\left[\ln W(\vec{\alpha})\right],
\end{equation}
where we have defined the expectation value of a function with respect to the probability distribution $W(\vec{\alpha})$ as $\mathbb{E}_{W}\left[\bullet\right] =\int\dd\vec{\alpha}W(\vec{\alpha})\bullet$. Using the definition of the Gaussian Wigner distribution in Eq.~\eqref{Eq: Wigner_Entropy_Expectation}, the Wigner entropy can be expressed as 
\begin{align}
	S(W) &= -\mathbb{E}_{W}\left[\ln(\pi^{-1})\right] - \mathbb{E}_{W}\left[\ln\left(|\Theta|^{-1/2}\right)\right] - \mathbb{E}_{W}\left[\ln\left(\exp\left(-\frac{1}{2}(\vec{\alpha}-\vec{v})^\dagger\Theta^{-1}(\vec{\alpha}-\vec{v})\right)\right)\right]\nonumber\\
	&= -\ln(\pi^{-1}) - \ln\left(|\Theta|^{-1/2}\right) -\mathbb{E}_{W}\left[\left(-\frac{1}{2}(\vec{\alpha}-\vec{v})^\dagger\Theta^{-1}(\vec{\alpha}-\vec{v})\right)\right] \nonumber\\
	&= \ln\left(\pi\right) +\frac{1}{2} \ln\left(|\Theta|\right) +\frac{1}{2}\mathbb{E}_{W}\left[(\vec{\alpha}-\vec{v})^\dagger\Theta^{-1}(\vec{\alpha}-\vec{v})\right].\label{Eq:Wig_Ent_break}
\end{align}
To tackle the final term of the right hand side of Eq. \eqref{Eq:Wig_Ent_break} we use the following trick. Let us consider a orthogonal basis $\vec{n}$ for a space of dim-2 satisfying $\sum_n \vec{n}\vec{n}^\dagger=\mathbb{I}$. Then, we can write 
\begin{align}
	\mathbb{E}_{W}\left[(\vec{\alpha}-\vec{v})^\dagger\Theta^{-1}(\vec{\alpha}-\vec{v})\right] &= \mathbb{E}_{W}\left[\sum_n(\vec{\alpha}-\vec{v})^\dagger\vec{n}\vec{n}^\dagger\Theta^{-1}(\vec{\alpha}-\vec{v})\right]\nonumber\\
	&= \mathbb{E}_{W}\left[\text{Tr}\left\{\Theta^{-1}(\vec{\alpha}-\vec{v})(\vec{\alpha}-\vec{v})^\dagger\right\}\right]\nonumber\\
	&= \text{Tr}\left\{\Theta^{-1}\mathbb{E}_{W}\left[(\vec{\alpha}-\vec{v})(\vec{\alpha}-\vec{v})^\dagger\right]\right\}\nonumber\\
	&= \text{Tr}\left\{\Theta^{-1}\Theta\right\} = 2,
\end{align}
where we have used the definition of the covariance matrix, $\Theta = \mathbb{E}_{W}\left[(\vec{\alpha}-\vec{v})(\vec{\alpha}-\vec{v})^\dagger\right]$. Inserting this expression into Eq.~\eqref{Eq:Wig_Ent_break}, the Wigner entropy for a Gaussian state is expressed as
\begin{equation}
	S(W) = \left[\ln(\pi)+1\right]+\frac{1}{2}\ln(|\Theta|).\label{wignentropy}
\end{equation}
A general expression for the Wigner relative entropy can be calculated in a similar manner. Given two Gaussian Wigner distributions, $W_1(\vec{\alpha})$ and $W_2(\vec{\alpha})$, with mean vectors and covariance matrices $\vec{v}_1$, $\vec{v}_2$ and $\Theta_1$, $\Theta_2$ respectively, the relative entropy is given by
\begin{equation}
	\label{Eq: KL_Gaussian}
	K\left[W_1||W_2\right] = \int\text{d}\vec{\alpha}W_1(\vec{\alpha})\ln\frac{W_1(\vec{\alpha})}{W_2(\vec{\alpha})} =  -S(W_1)-\mathbb{E}_{W_1}\left[\ln(W_2(\vec{\alpha}))\right].
\end{equation}
Then, by rewriting the second term on the right hand side as
\begin{align}
	\mathbb{E}_{W_1}\left[\ln(W_2(\vec{\alpha}))\right] &= \mathbb{E}_{W_1}\left[\ln(\pi^{-1}) + \ln\left(|\Theta_2|^{-1/2}\right) + \ln\left(\exp\left(-\frac{1}{2}(\vec{\alpha}-\vec{v}_2)^\dagger\Theta_2^{-1}(\vec{\alpha}-\vec{v}_2)\right)\right)\right]\nonumber\\
	&=-\ln(\pi) -\frac{1}{2}\ln\left(|\Theta_2|\right)-\frac{1}{2}\mathbb{E}_{W_1}\left[(\vec{\alpha}-\vec{v}_2)^\dagger\Theta_2^{-1}(\vec{\alpha}-\vec{v}_2)\right],\label{step2}
\end{align}
we can again use the cyclic property of the trace to calculate the last term on Eq. \eqref{step2}
\begin{align}
	\label{Eq:Expec_diff}
	\mathbb{E}_{W_1}\left[(\vec{\alpha}-\vec{v}_2)^\dagger\Theta_2^{-1}(\vec{\alpha}-\vec{v}_2)\right] = \text{Tr}\left\{\Theta_2^{-1}\mathbb{E}_{W_1}\left[(\vec{\alpha}-\vec{v}_2)(\vec{\alpha}-\vec{v}_2)^\dagger\right]\right\}.
\end{align}
Multiplying the terms in the expectation value, one finds
\begin{subequations}
	\begin{gather}
		\mathbb{E}_{W_1}\left[\vec{\alpha}\vec{\alpha}^{\dagger}\right] = \Theta_1 + \vec{v}_1\vec{v}_1^\dagger,\\
		\mathbb{E}_{W_1}\left[\vec{\alpha}\vec{v}^\dagger_2\right] = \vec{v}_1\vec{v}_2^\dagger,\\
		\mathbb{E}_{W_1}\left[\vec{v}_2\vec{\alpha}^\dagger\right] = \vec{v}_2\vec{v}_1^\dagger,\\
		\mathbb{E}_{W_1}\left[\vec{v}_2\vec{v}_2^\dagger\right] = \vec{v}_2\vec{v}_2^\dagger.
	\end{gather}
\end{subequations}
Plugging these identities into Eq.~\eqref{Eq:Expec_diff} one gets
\begin{equation}
	\mathbb{E}_{W_1}\left[(\vec{\alpha}-\vec{v}_2)^T\Theta_2^{-1}(\vec{\alpha}-\vec{v}_2)\right] = \text{Tr}\left\{\Theta_2^{-1}\Theta_1\right\} + (\vec{v}_1-\vec{v}_2)^\dagger\Theta_2^{-1}(\vec{v}_1-\vec{v}_2).
\end{equation}
Finally, combining everything we get
\begin{equation}
	\int \text{d} \vec{\alpha} W_1\ln(W_2) = -\ln(\pi) -\frac{1}{2}\ln\left(|\Theta_2|\right)-\frac{1}{2}\left(\text{Tr}\left\{\Theta_2^{-1}\Theta_1\right\} + (\vec{v}_1-\vec{v}_2)^\dagger\Theta_2^{-1}(\vec{v}_1-\vec{v}_2)\right).\label{step3}
\end{equation}
Then, using Eqs. \eqref{wignentropy} and \eqref{step3} in \eqref{Eq: KL_Gaussian}, a general expression for the Wigner relative entropy between Gaussian states is obtained:
\begin{equation}
	K[W_1||W_2] = -1 + \frac{1}{2}\left( \ln\left(\frac{|\Theta_2|}{|\Theta_1|}\right) + \text{Tr}\left\{\Theta_2^{-1}\Theta_1 \right\} + \left(\vec{v}_1-\vec{v}_2\right)^\dagger\Theta_2^{-1}\left(\vec{v}_1-\vec{v}_2\right)\right).
\end{equation}

\section{Ergotropy from the Wigner Function}
\label{AppaB}
In this appendix, we derive the expression for the ergotropy of Gaussian states [Eq. \eqref{werg4}]. We start computing the Wigner relative entropy between  an arbitrary state $W_1(\vec{a})$ and a thermal state, $W_{\rm th}(\vec{\alpha})$, which follows
\begin{align}
	K[W_1||W_{\rm th}]&=-\int \text{d} \vec{\alpha} W_1(\vec{\alpha}) \ln W_{\rm th}(\vec{\alpha})-S(W_1)+S(W_{\rm th})-S(W_{\rm th})\nonumber\\
	&=-\int \text{d} \vec{\alpha} W_1(\vec{\alpha}) \ln W_{\rm th}(\vec{\alpha})+\int d\vec{\alpha} W_{\rm th}(\vec{\alpha}) \ln W_{\rm th}(\vec{\alpha})+S(W_{\rm th})-S(W_1)\nonumber\\
	&=-\int \text{d} \vec{\alpha} (W_1(\vec{\alpha})-W_{\rm th}(\vec{\alpha})) \ln W_{\rm th}(\vec{\alpha})+S(W_{\rm th})-S(W_1).\label{Eq:wrelative2}
\end{align}
Here, we have added and subtracted the Wigner entropy of a thermal state with inverse temperature $\beta$ on the first line. By explicitly inserting the expression for the Wigner function of the thermal state, $W_{\rm th}(\vec{\alpha}) = \frac{1}{\pi f(\beta)}\exp(-\frac{|\alpha|^2}{f(\beta)})$, into Eq.~\eqref{Eq:wrelative2}, we obtain
\begin{align}
	K[W_1||W_{\rm th}]&=-\int \text{d} \vec{\alpha}(W_1(\vec{\alpha})-W_{\rm th}(\vec{\alpha})) \left(\ln\left(\frac{1}{\pi f(\beta)}\right) -\frac{|\alpha|^2}{ f(\beta)}\right)+S(W_{\rm th})-S(W_1)\nonumber\\
	&=\frac{1}{f(\beta)}\int \text{d} \vec{\alpha} (W_1(\vec{\alpha})-W_{\rm th}(\vec{\alpha}))|\alpha|^2+S(W_{\rm th})-S(W_1).\label{relatthw}
\end{align}
For a single-mode bosonic system, the Hamiltonian is defined as $\hat{H}=\omega\left(\hat{a}^\dagger\hat{a}+1/2\right)$. Rewriting this in symmetric ordering --- $\hat{H}=\omega/2\left(\hat{a}^\dagger\hat{a}+\hat{a}\hat{a}^\dagger\right)$ --- allows the calculation of the expectation energy through the Wigner function as
\begin{equation}
	E(W_1) = \omega\int\text{d} \vec{\alpha}\left|\alpha\right|^2 W_1(\vec{\alpha}),
\end{equation}
which can be inserted into the expression \eqref{relatthw} for the relative entropy 
\begin{equation}
	\label{Eq:KL_Gaussian_energy}
	K\left[W_1||W_{\rm th}\right] = \frac{1}{\omega f(\beta)}\left[E(W_1)-E(W_{\rm th})\right] +S(W_{\rm th})-S(W_1).
\end{equation}

Now, let us turn our attention to the ergotropy, which is defined by the energy difference between the non-passive state $W(\vec{\alpha})$ and its associated passive state $W_\pi(\vec{\alpha})$, i.e. 
\begin{equation}
	\mathcal{E}(W) = E(W)-E(W_\pi).\label{ergsm}
\end{equation}
By adding $E(W_{\rm th})-E(W_{\rm th})$ to the righthand side of Eq.~\eqref{ergsm} and then multiplying both sides by $(\omega f(\beta))^{-1}$, we obtain
\begin{align}
	\frac{1}{\omega f(\beta)}\mathcal{E}(W) = \frac{1}{\omega f(\beta)}[E(W)-E(W_{\rm th})]-\frac{1}{\omega f(\beta)}[E(W_\pi)-E(W_{\rm th})].\label{ergsm2}
\end{align}
Then, using the result \eqref{Eq:KL_Gaussian_energy} in \eqref{ergsm2}, we can rewrite \eqref{ergsm2} as
\begin{equation}
	\label{Eq. General Ergotropy}
	\mathcal{E}(W)= \omega f(\beta)\left(K\left[W||W_{\rm th}\right]-K\left[W_{\pi}||W_{\rm th}\right]+S(W)-S(W_\pi)\right).
\end{equation}
Then, as $W(\vec{\alpha})$ and its passive $W_\pi(\vec{\alpha})$ represent states that are unitarily connected, they must have the same entropy, i.e.,  $S(W)=S(W_\pi)$. Adding to that, as we are dealing with Gaussian states, it turns out that the passive state is actually a thermal state, this means that we can choose a temperature $\beta\rightarrow\beta_\pi$ such that $W_{\rm th}(\vec{\alpha})=W_{\pi}(\vec{\alpha})$ and $K[W_{\rm \pi}||W_{\rm th}]=0$. Finally, from the above discussion, we can simplify \eqref{Eq. General Ergotropy}, and the ergotropy for Gaussian states can be cast in the elegant form
\begin{equation}
	\mathcal{E}(W)= \omega f(\beta_\pi)K\left[W||W_{\pi}\right],
\end{equation}
which is the Eq. \eqref{werg4} of the main text. 

Now, using \eqref{wignentropy}, we can easily check that if two Wigner functions $W_1(\vec{\alpha})$ and $W_2(\vec{\alpha})$  satisfy $S(W_1)=S(W_2)$, then $|\Theta_1|=|\Theta_2|$. For any thermal state $W_{\rm th}(\vec{\alpha})$ defined by the inverse temperature $\beta$, we have that $f(\beta)=\sqrt{|\Theta|_{\rm th}}$. Then, as $W(\vec{\alpha})$ and its passive $W_\pi(\vec{\alpha})$ satisfy $S(W)=S(W_\pi)$, we obtain the constraint [Eq. (7) in the main text]
\begin{align}
	f(\beta_\pi)=\sqrt{|\Theta|},
\end{align}
where $\beta_\pi$ is the inverse temperature of the passive state $W_\pi(\vec{\alpha})$ and $\Theta$ the covariance matrix associated to the nonpassive state $W(\vec{\alpha})$.

\section{Separating the Ergotropy into Mean Vector  and Covariance Contributions}\label{AppaC}
In this appendix, we will separate the ergotropy into a component depending on the mean vector and a separate component depending solely on the covariance matrix. From the previous appendix, the ergotropy of a Gaussian state is given by
\begin{equation}
	\label{Eq: App_3_Ergotropy}
	\mathcal{E}(W)= \omega f(\beta_\pi)K\left[W||W_{\pi}\right].
\end{equation}
Expanding the relative entropy using the result in Appendix~\ref{AppA}, Eq. \eqref{Eq: App_3_Ergotropy} becomes
\begin{equation}
	\label{Eq: App_3_Ergotropy_modified}
	\mathcal{E}(W)= \omega f(\beta_\pi)\left[-1 + \frac{1}{2}\left( \ln\left(\frac{|\Theta_\pi|}{|\Theta|}\right) + \text{Tr}\left\{\Theta_\pi^{-1}\Theta\right\} + \left(\vec{v}-\vec{v}_\pi\right)^\dagger\Theta_\pi^{-1}\left(\vec{v}-\vec{v}_\pi\right)\right)\right].
\end{equation}
The mean vector for any thermal state is null, then $\vec{v}_\pi=\vec{0}$ and, as we discussed in the previous section, $|\Theta|=|\Theta_\pi|$. Substituting these terms into the expression \eqref{Eq: App_3_Ergotropy_modified}, we obtain
\begin{equation}
	\mathcal{E}(W)= \omega f(\beta_\pi)\left[-1 + \frac{1}{2}\left(\text{Tr}\left\{\Theta_\pi^{-1}\Theta\right\} + \vec{v}^\dagger\Theta_\pi^{-1}\vec{v}\right)\right].
\end{equation}
Now, $\pi$ is a thermal state, then the covariance matrix of $W_\pi(\vec{\alpha})$ is $\Theta_\pi=f(\beta_\pi)\mathbb{I}=\sqrt{|\Theta|}\mathbb{I}$, allowing us to write 
\begin{equation}
	\mathcal{E}(W) =\omega |\vec{v}|^2+\omega \sqrt{|\Theta|}\left[\frac{1}{2 \sqrt{|\Theta|}}{\rm Tr}\{\Theta\}-1\right].
\end{equation}
We see that the ergotropy can then be split as $\mathcal{E}(W)=\mathcal{E}_{\rm d}(\vec{v})+\mathcal{E}_{\rm s}(\Theta)$,
\begin{align}
	&\mathcal{E}_{\rm d}(\vec{v})=\omega |\vec{v}|^2,\\
	&\mathcal{E}_{\rm s}(\Theta)=\omega \sqrt{|\Theta|}\left[\frac{1}{2 \sqrt{|\Theta|}}{\rm Tr}\{\Theta\}-1\right],
\end{align}
where $\mathcal{E}_{\rm d}(\vec{v})$ is the contribution arising due to the mean vector of the Gaussian state and $\mathcal{E}_{\rm s}(\Theta)$ is the contribution due to the covariance matrix of the state.

\section{Derivation and Analytical Solution of the Lyapunov Equation}\label{AppaD}
In this appendix, we calculate analytic solutions of the Lyapunov equation [Eq. \eqref{lyapunov}].  By taking the time derivative of the covariance matrix $\Theta$, Eq. \eqref{covmatrix}, we find that
\begin{subequations}
	\begin{align}
		&\dot{\Theta}_{11}=\dot{\Theta}_{22}=\frac{d\moy{\hat{a}^\dagger \hat{a}}}{dt}-\moy{\hat{a}}\frac{d \moy{\hat{a}^\dagger}}{dt}-\moy{\hat{a}^\dagger}\frac{d \moy{\hat{a}}}{dt},\label{dtheta11}\\
		&\dot{\Theta}_{12}=\dot{\Theta}^{*}_{21}=\frac{d \moy{\hat{a}\hat{a}}}{dt}-2\moy{\hat{a}}\frac{d \moy{a}}{dt},\label{dtheta12}
	\end{align}
\end{subequations}
where we are omitting the time dependent argument to make the notation clear. We explicitly denote the time dependence only when necessary. Using the master equation [Eq. \eqref{MEQ}] we can compute the derivative of the mean values on Eqs. \eqref{dtheta11} and \eqref{dtheta12}, which are given by
\begin{subequations}
	\begin{align}
		&\frac{d \langle \hat{a}\rangle}{dt}=\left(\frac{d \langle\hat{a}^\dagger\rangle}{dt}\right)^\dagger=-(\gamma/2+i \omega)\langle \hat{a}\rangle,\\
		&\frac{d \langle \hat{a}^\dagger \hat{a}\rangle}{dt}=\frac{d\langle \hat{a}\hat{a}^\dagger\rangle}{dt}=\gamma(\bar{n}-\langle \hat{a}^\dagger \hat{a}\rangle),\\
		&\frac{d \langle \hat{a} \hat{a}\rangle}{dt}=\left(\frac{d \langle \hat{a}^\dagger \hat{a}^\dagger\rangle}{dt}\right)^\dagger=-(\gamma+2i \omega)\langle \hat{a}\hat{a}\rangle.
	\end{align}
\end{subequations}
Now, plugging these results in the equations of motion for $\Theta$, Eqs. \eqref{dtheta11} and \eqref{dtheta12}, we get
\begin{subequations}
	\begin{align}
		&\dot{\Theta}_{11} = \dot{\Theta}_{22}=-\gamma( f(\beta)+\Theta_{11}),\\
		&\dot{\Theta}_{12}=\dot{\Theta}^{*}_{21} = -\left(\gamma-2i\omega\right)\Theta_{12},
	\end{align}
\end{subequations}
where we see that the diagonal and out-of-diagonal elements of $\Theta$ are not coupled. This equation is known as Lyapunov Equation, and can be cast in its familiar  form as
\begin{align}
	&\dot{\Theta}=\Lambda \Theta+\Theta \Lambda^\dagger + F,\\
	&\Lambda=-\frac{1}{2}\begin{bmatrix}
		\gamma+2i\omega & 0 \\
		0 & \gamma-2i\omega 
	\end{bmatrix}, \ \ 
	F=f(\beta)\hat{\mathbb{I}}. \nonumber
\end{align}
The solution for this Lyapunov equation can be found using simple integration techniques, and is given by
\begin{subequations}
	\begin{align}
		&\Theta_{11}(t)=c_{11}^0 e^{-\gamma t}+f(\beta),\\
		&\Theta_{12}(t)=c_{12}^0e^{-(\gamma+2i\omega)t},
	\end{align}
	where the constants $c_{ij}^0$ are obtained from the initial conditions.
\end{subequations}

\section{Ergotropy Contributions from Displaced and Squeezed Thermal States}\label{AppaE}
In this appendix, we are going to show that the contributions $\mathcal{E}_{\rm d}(\vec{v})$ and $\mathcal{E}_{\rm s}(\Theta)$ depend only on the displacement and squeezing parameters, respectively. First, let us obtain the covariance matrix and mean vector for the system initially prepared in a displaced thermal state $\hat{\rho}_{\rm d}=\hat{D}(\mu)\hat{\pi} \hat{D}^\dagger(\mu)$, with $\hat{D}(\mu)=\exp[\mu \hat{a}^\dagger-\mu^*\hat{a}]$. Using the following properties of the displacement operator \cite{PKnight}
\begin{align}
	&\hat{D}^\dagger(\mu)\hat{a}\hat{D}(\mu)=\hat{a}+\mu,\\
	&\hat{D}^\dagger(\mu)\hat{a}^\dagger \hat{D}(\mu)=\hat{a}^\dagger+\mu^*,
\end{align}
we have that $\vec{v}(t)=(\mu_t,\mu_t^*)$, with $\mu_t=\mu e^{-(i\omega+\gamma/2)t}$. We can also compute the covariance matrix, which is given by $\Theta(t)=\Delta_\beta \hat{\mathbb{I}}$, where we defined the $\Delta_\beta=[f(\beta_\pi)-f(\beta)]e^{-\gamma t}+f(\beta)$. 
Here, we note some important results. First, the displacement operator only affects the mean vector $\vec{v}$. This means that $\Theta$ is the same as the covariance matrix that would be obtained if we solved the Master Equation [Eq. \eqref{MEQ}] for the initial state being the thermal passive state $\hat{\pi}$. From the mean vector and covariance matrix dynamics, we can compute the time dependent Wigner function for the displaced thermal state, 
\begin{align}
	W_{\rm d}(\vec{\alpha})=\frac{1}{\pi \Delta_\beta }e^{-\frac{|\alpha-\mu_t|^2}{\Delta_\beta }},\label{wdeq} 
\end{align}
which is a Gaussian displaced by $\mu_t$. From the Wigner function and covariance matrix, we can compute the time dependence of the energy of the displaced state and its passive state,
\begin{align}
	&E(W_{\rm d})=\omega \Delta_\beta+\omega|\mu_t|^2,\\
	&E(W_{\pi_{\rm d}})=\omega \Delta_\beta.\label{passived}
\end{align}
Then, the ergotropy is given by
\begin{align}
	\mathcal{E}(W_d)=\omega|\mu_t|^2, 
\end{align}
where we see that the thermal contribution of the energy $E(W_{\rm d})$ cancels out with the energy of the passive state, and as a consequence the ergotropy for a displaced thermal state only depends on the displacement parameter $\mu$. It is worth remarking that as the displacement operator does not affect the covariance matrix, we have that $\mathcal{E}_{\rm s}(\Theta)=0$ and the total ergotropy for the displaced thermal state is completely determined by the mean vector contribution, i.e, $\mathcal{E}(W_{\rm d})=\mathcal{E}_{\rm d}(\vec{v})$.

Now, let us look at the solution of the Lyapunov equation when the initial state is the squeezed thermal state $\hat{\rho}_{\rm s}=\hat{S}(z)\hat{\pi}\hat{S}^\dagger(z)$, where $\hat{S}(z)=\exp[\frac{1}{2}(z^*\hat{a}^2-z\hat{a}^{\dagger2})]$ and $z=re^{i\theta}$. The squeezing operator satisfies the following mode transformations \cite{PKnight}
\begin{align}
	&\hat{S}^{\dagger}(z)\hat{a}\hat{S}(z)=\hat{a}\cosh(r)-\hat{a}^\dagger e^{i\theta}\sinh(r),\\
	&\hat{S}^{\dagger}(z)\hat{a}^{\dagger}\hat{S}(z)=\hat{a}^\dagger\cosh(r)-\hat{a} e^{-i\theta}\sinh(r).
\end{align}
For the squeezed thermal state we have that $\vec{v}(t)=\vec{0}$, in other words, the squeezing operator does not affect the mean vector. However, the squeezing operator does affect the covariance matrix, whose the elements in this case are given by
\begin{align}
	&\Theta_{11}(t)=(f(\beta_\pi)\cosh(2r)-f(\beta))e^{-\gamma t}+f(\beta),\\
	&\Theta_{12}(t)=-e^{i \theta_t}f(\beta_\pi)\sinh(2r)e^{-\gamma t},
\end{align}
where $\theta_t=\theta-2\omega t$. From the mean vector and covariance matrix, after some algebra the time dependent Wigner function for a squeezed thermal state can be written as
\begin{align}
	W_{\rm s}(\vec{a})=\frac{1}{\pi f(\beta_t)}\exp\left[\frac{-|\alpha \cosh(r_t)+e^{i\theta_t}\alpha^*\sinh(r_t)|^2}{{f(\beta_t)}}\right],
\end{align}
where $r_t$ and $\beta_t$ are implicitly defined by the functions
\begin{align}
	& f(\beta_t)=\sqrt{\Delta_\beta^2+4f(\beta_\pi)f(\beta)e^{-2\gamma t}(e^{\gamma t}-1)\sinh^2(r)},\\
	& \cosh(2 r_t)=\frac{1}{f(\beta_t)}[\Delta_\beta+2f(\beta_\pi)e^{-\gamma t}\sinh^2(r)].
\end{align}
We can then compute the energy of the squeezed thermal state and its passive state, which are given by
\begin{align}
	&E(W_{\rm s})=\omega f(\beta_t) \cosh(2 r_t),\\
	&E(W_{\pi_{\rm s}})=\omega f(\beta_t),
\end{align}
and the ergotropy
\begin{align}
	\mathcal{E}(W_{\rm s})=\omega f(\beta_t) [\cosh(2 r_t)-1].
\end{align}
Here, we notice a fundamental difference from the displaced thermal state. As the squeezing operation affects the thermal contribution to the energy of the state, which is multiplied by a factor $\cosh(2 r_t)$, the ergotropy will have a dependence on $\beta$ and $\beta_{\pi}$. Moreover, the temperature of the passive state is also affected by the squeezing parameter in such a way that its evolution is completely different from the passive state associated to a displaced thermal state (Eq. \eqref{passived}). We see that $E(W_{\pi_{\rm s}})\geq E(W_{\pi_{\rm d}})$ due to the positive contribution that comes from the squeezing parameter in the passive state energy. As discussed in the main text, this difference in the energy of the passive states is responsible for the ergotropic Mpemba effect for Gaussian states. 

Finally, we have seen that $\hat{D}(\mu)$ only affects $\vec{v}$ and $\hat{S}(z)$ only affects $\Theta$. In addition, an arbitrary Gaussian state can be generated as $\hat{\rho}=\hat{S}(z)\hat{D}(\mu) \hat{\pi}\hat{D}^\dagger(\mu)\hat{S}^\dagger(z)$ and $\Theta$ and $\vec{v}$ evolve independently of each other. Then, $\mathcal{E}_{\rm d}(\vec{v})=\mathcal{E}(W_{\rm d})=\mathcal{E}_{\rm d}(\mu)$ will depend only on $\hat{D}(\mu)$ and $\mathcal{E}_{\rm s}(\Theta)=\mathcal{E}(W_{\rm s})=\mathcal{E}_{\rm s}(r)$ will depend only on $\hat{S}(z)$, allowing us to write
\begin{align}
	&\mathcal{E}_{\rm d}(\mu)=\omega |\mu|^2,\\
	&\mathcal{E}_{\rm s}(r)=\omega f(\beta_\pi)[\cosh(2 r)-1],
\end{align}
with the total ergotropy for an arbitrary single mode Gaussian state $\hat{\rho}$ being $\mathcal{E}(W)=\mathcal{E}_{\rm d}(\mu)+\mathcal{E}_{\rm s}(r)$.

\end{document}